\documentclass[12pt,a4paper,reqno]{article}

\usepackage{amsmath,latexsym,amssymb,amsthm}
\usepackage{graphicx}
\usepackage[hypertex]{hyperref}

\usepackage{amsmath}\usepackage{amssymb}
\usepackage{amsthm}

\newcommand{\semilinespace}{\vspace{0.5\baselineskip}}
\newcommand{\upline}{\vspace{-\baselineskip}}

\newcommand{\be}{\begin{equation}}
\newcommand{\ee}{\end{equation}}

\newcommand{\e}{\mathrm{e}}

\renewcommand{\H}{\mathcal{H}}

\newcommand{\ox}{\otimes}

\renewcommand{\>}{\rangle}
\newcommand{\half}{\tfrac{1}{2}}

\newcommand{\third}{\tfrac{1}{3}}

\newcommand{\head}{\text{head}}
\newcommand{\tail}{\text{tail}}
\newcommand{\ok}{\text{ok}}
\newcommand{\fail}{\text{fail}}
\theoremstyle{definition}

\theoremstyle{remark}

\numberwithin{equation}{section}

\begin{document}

\title{\bf{The hidden assumptions of Frauchiger and Renner}}

\author{Anthony Sudbery$^1$\\[10pt] \small Department of Mathematics,
University of York, \\[-2pt] \small Heslington, York, England YO10 5DD\\
\small $^1$ tony.sudbery@york.ac.uk}

\date{}

\maketitle

\begin{abstract}

This note is a critical examination of the argument of Frauchiger and Renner \cite{FR2}, in which they claim to show that three reasonable assumptions about the use of quantum mechanics jointly lead to a contradiction. It is shown that further assumptions are needed to establish the contradiction, and that each of these assumptions is invalid in some version of quantum mechanics.

\end{abstract}

\section{Introduction}

In the title of \cite{FR2} Frauchiger and Renner (hereafter FR) proclaim that ``Quantum mechanics cannot consistently describe the use of itself". To justify this, they claim to prove that a certain set of three assumptions, labelled (Q), (C) and (S), cannot be jointly satisfied in quantum mechanics. I refer to \cite{FR2} for their own precise statements of these assumptions; they can be expressed as follows.

\begin{quote}{\bf Assumption (Q)}\ \  If an agent $A$ is certain that a system is in an eigenstate of an observable $X$ with eigenvalue $\xi$  at time $t_0$, and a measurement of $X$ is completed at time $t > t_0$, then $A$ is certain that the result $x$ of the measurement is $x = \xi$ at time $t$.\end{quote}

\begin{quote}{\bf Assumption (C)}\ \  If agent $A$ is certain that another agent $A'$ is certain that $x = \xi$ at time $t$, on the same grounds as in (Q), then $A$ is also certain that $x = \xi$ at time $t$.\end{quote}

\begin{quote}{\bf Assumption (S)}\ \  If $A$ is certain that the proposition $x = \xi$ is true, then $A$ is certain that the proposition $x \ne \xi$ is not true.\end{quote}

Assumption (C) is the property of consistency which Frauchiger and Renner seek to deny to quantum mechanics.  These assumptions will be discussed in section \ref{assumptions}.

The argument proceeds by applying these assumptions in analysing a Gedankenexperiment obtained by extending the ``Wigner's friend'' experiment described by Wigner in \cite{Wigner:consciousness}. This Gedankenexperiment is described in Section \ref{experiment}. The argument of FR is laid out in steps in Section \ref{argument}; at each step we consider whether it follows from the assumptions declared by FR, and, if not, what further assumptions are needed to establish it. In Section \ref{assumptions} we examine each of the assumptions and discuss whether they are held to be true in various versions (or interpretations) of quantum mechanics.

\section{The Gedankenexperiment}\label{experiment} 
The experiment which is supposed to demonstrate that the assumptions (Q), (C) and (S) lead to a contradiction contains four systems $F_1, F_2, W_1$ and $W_2$. These systems represent agents. Each agent $X$ has three orthogonal states $|0\>_X, |a\>_X$ and $|b\>_X$ where $|0\>_X$ is a state of readiness before $X$ participates in the experiment and $|a\>$ and $|b\>$ are states in which $X$ is certain that the result of a certain observable $s_X$ is given by the label $a$ or $b$. In addition, there are two qubit systems: a coin $C$ and the spin state $S$ of an electron, with orthogonal states $|\text{head}\>_C$ and $|\text{tail}\>_C$ for $C$, $|\pm\>_S$ for $S$. I will refer to $F_1$ and $F_2$, who operate on the physical systems $C$ and $S$, as ``experimenters", and $W_1$ and $W_2$ as ``para-experimenters" since they operate on the experimenters $F_1$ and $F_2$. 
%The agents $F_1$ and $W_1$ are supposed to be isolated from $F_2$ and $W_2$ until the end of the experiment, when $W_1$ and $W_2$ compare notes, but the system $S$ is transported from $F_1$ to $F_2$ in the course of the experiment. 

The experiment, the extended Wigner's friend experiment or EWF, proceeds in several stages. 
%At each step I will write down the state of the combined system $C\ox S\ox F_1\ox F_2\ox A\ox W$ which follows from quantum evolution according to the Schr\"odinger equation, with no further assumptions.%

\semilinespace

{\bf Stage EWF($-1$})\ \  Before the experiment starts the coin is prepared in the state 
$\sqrt{\third}|\head\> + \sqrt{\tfrac{2}{3}}|\tail\>$.

\semilinespace 

{\bf Stage EWF0}\ \  At time $t = 0$ experimenter $F_1$ observes the coin and records the result $r =$ ``head" or ``tail", thereby being put into a memory state $|r\>_{F_1}$.

\semilinespace

{\bf Stage EWF1}\ \  At time $t = 1$, $F_1$ prepares the electron as follows: if the result of the measurement at $t = 0$ was $r =$ ``head", $F_1$ prepares the electron in spin state $|\downarrow\>_S$; if $r = $ ``tail", they prepare it in spin state 

\[
|\rightarrow\>_S = \tfrac{1}{\sqrt{2}}\big(|\uparrow\>_S + |\downarrow\>_S.
\]

{\bf Stage EWF2}\ \  At time $t = 2$ experimenter $F_2$ measures the spin $z\half\hbar$ of the electron $(z = \pm)$ in the basis $\{|\uparrow\>, |\downarrow\>\}$ and records the result, thereby being put into a memory state $|z\>_{F_2}$.

\semilinespace

{\bf Stage EWF3}\ \  At time $t = 3$ para-experimenter $W_1$ measures $F_1$, together with the coin, in the basis
\begin{align*}
|\ok\>_{F_1C} &= \tfrac{1}{\sqrt{2}}\big(|\head\>_{F_1}|\head\>_C - |\tail\>_{F_1}|\tail\>_C\big)\\
|\fail\>_{F_1C} &= \tfrac{1}{\sqrt{2}}\big(|\head\>_{F_1}|\head\>_C + |\tail\>_{F_1}|\tail\>_C\big),
\end{align*}
and records the result $w_1 =$ ``ok" or ``fail".

\semilinespace

{\bf Stage EWF4}\ \  At time $t = 4$ para-experimenter $W_2$ measures $F_2$, together with the electron, in the basis
\begin{align*}
|\ok\>_{F_2S} &= \tfrac{1}{\sqrt{2}}\big(|-\>_{F_2}|\downarrow\>_S - |+\>_{F_2}|\uparrow\>_S\big)\\
|\fail\>_{F_2S} &= \tfrac{1}{\sqrt{2}}\big(|-\>_{F_2}|\downarrow\>_S + |+\>_{F_1}|\uparrow\>_S\big),
\end{align*}
and records the result $w_2 =$ ``ok" or ``fail".

\semilinespace

{\bf Stage EWF5}\ \  At the end of the experiment the two para-experimenters compare the results of their measurements. The question is whether it is possible for $w_1 = w_2 = $ ``ok".

\semilinespace

\section{The Argument}\label{argument}

Frauchiger and Renner argue as follows.

\semilinespace 

{\bf Step FR1.}\ \  Suppose that $F_1$ got the value $r = $ ``tails" in the measurement on the system $C$. Then $F_1$ will prepare the system $S$ in the state $|\rightarrow\>$. After this, therefore, let us say at  time $t = 1.5$, $F_1$ is certain that $S$ is in the state $|\rightarrow\>$.

This deduction requires the undeclared assumption

\begin{quote} {\bf Assumption (P)}\ \  If an agent $A$ prepares a system $S$ in a state $|\psi\>$, then immediately after the preparation $A$ is certain that $S$ is in the state $|\psi\>$.\end{quote}

What could be more innocent? Surely this needs stating even less than the logical assumption (S). I do not believe so, but will defer discussion until the next section.

\semilinespace

{\bf Step FR2.}\ \  $F_1$ proceeds from the certainty that $S$ is in the state $|\rightarrow\>$ at time $t = 1.5$ to the certainty that after time $t = 2$, when the electron $S$ has been sent to $F_2$ and measured by $F_2$, the state of $F_2$ and $S$ (the lab $L$ in \cite{FR2}) will be $\sqrt{\tfrac{1}{2}}\big(|-\>_{F_2 S} + |+\>_{F_2 S}\big)$ and therefore orthogonal to $|\ok\>_{F_2 S}$.

This step requires a stronger assumption than (Q): the certainty of $F_!$ about future events comes from reliance on the predictions of quantum mechanics applied to the whole world external to $F_1$. In particular, $F_1$ must treat agent $F_2$ as part of a quantum system, whose states, therefore, may not have definite values. This is

\begin{quote} {\bf Assumption (U)}\ \  If an agent $A$ is certain at time $t$ that the external world is in a state $|\Psi\>$ and $A$ knows the Hamiltonian $H$ governing the external world, then $A$ is certain at time $t$ that the state of the external world at time $t + T$ will be $\e^{-iHT/\hbar}|\Psi\>$.\end{quote}

The need for this assumption has also been noted by Nurgalieva and del Rio \cite{delRio:modal}.

\semilinespace

{\bf Step FR3}\ \  It follows from this, by Assumption (Q), that the measurement of $F_2S$ by $W_2$ at time $t = 3$ will give the result $w =$ ``fail''. Thus

\begin{quote}If $F_1$ at time $t = 0$ gets the result $r =$ ``tails", then $F_1$ is certain at time $t = 1.5$ that $W_2$ will observe $w = $ ``fail'' at time $t = 3$. \end{quote}

{\bf Step FR4}\ \  Now suppose that the result of the measurement of $S$ by $F_2$ at time $t = 2$ is $z = +$. Then $F_2$ is certain after this measurement, let us say at time $t = 2.5$, that the state of $S$ when it was passed on by $F_1$ was not $|\downarrow\>$, and therefore that $F_1$ did not get the result $r = $ ``head" from the measurement of the coin $C$.

\semilinespace

{\bf Step FR5}\ \  It follows (to $F_2$), by Assumption (S), that $F_1$ got the result $r = $ ``tail".

This does not follow from Assumption (S). What FR are assuming appears to be the converse, 

\begin{quote}{\bf Assumption ($\overline{\text{S}}$)}\ \  If $A$ is certain that the proposition $x \ne \xi$ is not true, then $A$ is certain that $x = \xi$ is true, \end{quote}

\noindent but for use in this step of the argument, this assumption needs to be stated more precisely as

\begin{quote} {\bf Assumption (T)}\ \  If the possible values of a measurement result are $r_1, r_2, \ldots$ and an agent $A$ is certain that the result was not $r_1$, then $A$ is certain that the result was one of $r_2, \ldots$. \end{quote}

\semilinespace 

{\bf Step FR6}\ \  Therefore, by the conclusion of Step FR3, $F_2$ is certain at time $t = 2.5$ that $F_1$ was certain at time $t = 1.5$ that $W_2$ will observe $w = $ ``fail'' at time $t = 3$.

\semilinespace 

{\bf Step FR7}\ \  By Assumption (C), $F_2$ is certain at time $t = 2.5$ that $W_2$ will observe $w = $ ``fail'' at time $t = 3$.

This step requires consideration of the times at which the agents hold their various certainties. In the statement by FR, Assumption (C) contains no mention of these times. If it is assumed that it refers to certainties held at the same time, then the deduction FR5 assumes that the memory state of $F_1$ is unchanged between the times $t = 1.5$ and $t = 2.5$. This is reasonable, since there has apparently been no dynamical action on $F_1$ in this time interval, but it needs to be deduced from some general assumption. One possible form for this assumption is 

\begin{quote} {\bf Assumption (L)}\ \  If there is no action in the neighbourhood of a system $F$ in a certain time interval, then the state of $F$ remains unchanged during that interval.\end{quote}

An alternative assumption which would justify Step FR7 is

\begin{quote} {\bf Assumption (M)}\ \  Memory states of an agent are irreversible and do not change once they have been established unless there is some surgical operation on, or neural pathology in, the agent.\end{quote}

\noindent Note that we consider memory states to be the same as states of certainty.

\semilinespace

{\bf Step FR8}\ \  FR now consider the measurement of $F_1C$ by $W_1$ in Stage EWF3, in conjunction with the measurement of $S$ by $F_2$ in Stage EWF2. They argue that the eigenvectors of this joint measurement are
\[
|\ok,z\>_{F_1CS} = |\ok\>_{F_1C}|z\>_S \;\;\; \text{and} \;\;\; |\fail,z\>_{F_1CS} = |\fail\>_{F_1C}|z\>_S. 
\]
More precisely, taking account of the fact that when $W_1$ measures $F_1C$, the experimenter $F_2$ has measured $S$ and therefore $F_2$ and $S$ have become entangled, so that the measurement by $W_1$ is enacted on the space $\H_{F_1CS}\ox\H_{F_2}$, we should consider the eigenspaces of this measurement which are 
\[
\H_{\ok,z} = |\ok\>_{F_1C}|z\>_S\ox\H_{F_2}  \;\;\; \text{and} \;\;\; \H_{\fail,z} = |\fail,z\>_{F_1C}|z\>_S\ox\H_{F_2}.
\]
The measurement is performed on the state
\[
\frac{1}{\sqrt{6}}\big(2|\fail\>_{F_1C}|\downarrow\>_S|-\>_{F_2} + |\fail\>_{F_1C}|\uparrow\>_S|+\>_{F_2} - |\ok\>_{F_1C}|\uparrow\>_S|+\>_{F_2}\big)
\]
which is orthogonal to $\H_{\ok,-}$. It is therefore not possible that $W_1$ gets the result $w_1 =$ ``ok" while $F_2$ gets the result $z = -$.

\semilinespace

{\bf Step FR9}\ \  Hence if $W_1$ gets the result $w_1 = $ ``ok", then $W_1$ at time $t = 3.5$ is certain that $F_2$ must have got the result $z = +$ and therefore, by Step FR7, {$F_2$ was certain at time $t = 2.5$ that $W_2$ will get the result $w_2 = $ ``fail" at time $t = 5$.

\semilinespace

{\bf Step FR10}\ \  It follows by Assumptions (C) and (L) that $W_1$ is certain at time $t = 3.5$ that $W_2$ will get the result $w_2 = $ ``fail" at time $t = 5$.

\semilinespace

{\bf Step  FR11}\ \  When $W_1$ announces this certainty to $W_2$, the latter will also become certain, by Assumption (C), that the result of the final measurement must be $w_2 =$ ``fail". Thus

\begin{quote} If $W_1$ gets the result $w_1 = $ ``ok", $W_2$ is certain at time $t = 3.5$ that he will get the result $w_2 =$ ``fail". \end{quote}

{\bf Step FR12}\ \  But $W_2$ can calculate the probability that the results of the final two measurements will be $w_1 = w_2 = $ ``ok"; a straightforward quantum-mechanical calculation (e.g. \cite{singleworld}) shows that this probability is 1/12. $W_2$ is therefore certain that the result of the measurements of $W_1$ and $W_2$ can be $w_1 = w_2 = $``ok", in contradiction to the conclusion of Step FR11.

 \section{Discussion of the Assumptions}\label{assumptions}

\subsection{}\label{Q}
\upline
\begin{quote}{\bf Assumption (Q)}\ \  If an agent $A$ is certain that a system is in an eigenstate of an observable $X$ with eigenvalue $\xi$  at time $t_0$, and a measurement of $X$ is completed at time $t > t_0$, then $A$ is certain that the result $x$ of the measurement is $x = \xi$ at time $t$.\end{quote}

The wording ``$A$ is certain that $x = \xi$ at time $t$" is taken directly from \cite{FR2}. It is ambiguous, but must be understood to mean ``$A$ is certain at time $t$ that $x = \xi$". The alternative reading is that the content of $A$'s certainty is ``$x = \xi$ at time $t$".  But this does not make sense, since $x$ is a single value, the result of the experiment, and does not depend on $t$. 

With this understanding, Assumption (Q) appears to be a standard assertion of quantum mechanics. FR discuss whether it holds in all versions of quantum mechanics, and conclude that it does not hold in hidden-variable theories such as Bohmian mechanics, appealing to an unspecified calculation in Bohmian mechanics in which, they assert, the conclusion of Step FR3 does not hold. It is not clear what supports this assertion, since Bohmian mechanics requires particle positions which are not present in their scenario. The Gedankenexperiment can be analysed in Bell's probabilistic extension of Bohmian mechanics \cite{singleworld}. Here the hidden variables are the memories (or certainties) of the four agents, which have definite values at all times. The result of the Bell-Bohmian analysis is that a possible sequence of memories of the agents is as follows:

\semilinespace
 
\hspace{2em} At Stage EWF($-1$):  $|0\>_{F_1}|0\>_{F_2}|0\>_{W_1}|0\>_{W_2}$

\hspace{2em} After Stage EWF0: $|\tail\>_{F_1}|0\>_{F_2}|0\>_{W_1}|0\>_{W_2}$

\hspace{2em} After Stage EWF2: $|\tail\>_{F_1}|+\>_{F_2}|0\>_{W_1}|0\>_{W_2}$

\hspace{2em} After Stage EWF3: $|\tail\>_{F_1}|+\>_{F_2}|\ok\>_{W_1}|0\>_{W_2}$

\hspace{2em} After Stage EWF4:  $|\tail\>_{F_1}|-\>_{F_2}|\ok\>_{W_1}|\ok\>_{W_2}$.

\semilinespace

This model does not contain variables describing the beliefs of any of the agents about the other agents or other aspects of the external world. However, we can assume that in each of the states labelled as above, each agent has consulted the other agents and therefore is certain that the beliefs of the other agents are given by the labels in the state. Then at time $t = 1.5$ agent $F_1$ is certain that the state of $S$ is $|\rightarrow\>$, in agreement with Assumption (Q), but at time $t = 2.5$ they are certain that the state of $F_2$ is $|+\>_{F_2}$, so that Step FR2 is invalid. Thus it is not Assumption (Q) but (U) that may be violated in Bohmian mechanics. This is probably also true in the Copenhagen interpretation of quantum mechanics, which can be taken to assert that macroscopic variables such as the beliefs of agents always have definite values.

\semilinespace

%It will be seen that Assumption (Q) is not violated: after measuring the coin and getting the value ``tail", $F_1$ remains at all time certain that the result of the measurement was ``tail". The source of the avoidance of the contradiction in Bell-Bohmian theory must be found elsewhere.

\subsection{}\label{C} 
\upline
\begin{quote}{\bf Assumption (C)}\ \  If agent $A$ is certain that another agent $A'$ is certain that $x = \xi$ at time $t$, on the same grounds as in (Q), then $A$ is also certain that $x = \xi$ at time $t$.\end{quote}

I find it hard to see how this assumption is violated in any interpretation of quantum mechanics, except possibly QBism. FR assert that it is violated in five different interpretations, but they discuss only three of these. In two of these cases their argument is that Assumptions (Q) and (S) are satisfied and therefore Assumption (C) must be violated, otherwise their theorem would show that the interpretation contains a contradiction. As shown above, this inference is invalid; all that is shown is that one of the assumptions (C), (P), (U) or (T), or both of (L) and (M), are violated. 

In the third case, that of the Consistent Histories (CH) interpretation \cite{Griffiths:book}, FR offer to illustrate how the violation of (C) manifests itself. They consider the statements
\begin{quote} $h_1$: The outcomes $r =$ ``tails", $z = +$, $w_1 =$ ``ok" and $w_2 =$ ``ok" were observed" \end{quote}
and
\begin{quote} $h_1'$: The outcomes $r = $ ``tails'' and $w_2 =$ ``ok'' were observed \end{quote}
and calculate the probabilities
\[ P[h_1] = \frac{1}{12} \quad \text{and} \quad P[h_1'] = 0 \]
by considering $h_1$ and $h_1'$ as histories in the sense of the CH interpretation, namely as sequences of projectors. These probabilities appear to be in conflict with the fact that $h_1$ implies $h_1'$, but FR note that the CH interpretation does not allow these two histories to be considered together, and interpret this as a violation of Assumption (C). However, the meaning of the history (sequence of projectors) into which FR translate $h_1'$ is not $h_1'$ but
\begin{quote} $h_1''$: The outcomes $r = $ tails and $w_2 = ok$ were observed, and $z$ and $w_2$ were not observed, \end{quote}
which clearly is not implied by $h_1$. The tension between $h_1$ and $h_1''$ looks more like a violation of Assumption (S), which we discuss next.

\subsection{}\label{S}
\upline
\begin{quote}{\bf Assumption (S)}\ \  If agent $A$ is certain that the proposition $x = \xi$ is true, then $A$ is certain that the proposition $x \ne \xi$ is not true.\end{quote}

As stated here, this is satisfied in all versions of quantum mechanics. What may not be satisfied is the converse,

\begin{quote}{\bf Assumption ($\overline{\text{S}}$)}\ \  If $A$ is certain that the proposition $x \ne \xi$ is not true, then $A$ is certain that $x = \xi$ is true. \end{quote}

%(The difference between these two can be illustrated from everyday experience: If I am well, it is certainly true that I am not unwell; but if I am not unwell, I may not actually be well.) 
It is Assumption ($\overline{\text{S}}$) which FR use in Step FR5. However, there is still some ambiguity in the above Assumption ($\overline{\text{S}}$). Here (as in FR's statement of Assumption S), the nature of $x$ is unspecified. If $x$ is a number or other classical variable, then Assumption (S) is equivalent to ($\overline{\text{S}}$). %(My state of health is far from classical.)
But if $x$ is a quantum observable, then the proposition $x = \xi$ means that the state of the system lies in the eigenspace of $x$ with eigenvalue $\xi$, while $x\ne\xi$ means that it lies in the orthogonal subspace.  Clearly, then, $x = \xi$ implies the denial of $x\ne\xi$, but the converse implication does not hold. 

In FR's argument the status of $x$ shifts in the course of the argument. Originally, $x$ is defined to be the result of a measurement (by $F_1$), and in standard quantum mechanics this is a classical variable. In the later stages of the Gedankenexperiment, however, $x$, the content of $F_1$'s memory, is a quantum variable and therefore Assumption ($\overline{\text{S}}$) is invalid. In Step FR5 $F_2$ argues as if Assumption (T) refers to a classical variable; but knowing that $W_1$ is going to operate on $F_1$ as a quantum system, $F_1$'s deduction in Step 5 should proceed via Assumption ($\overline{\text{S}}$) with $x$ understood to be a quantum variable.

In their table 4 \cite{FR2} FR assert that Assumption (S) is not satisfied in the ``many worlds'' interpretation of quantum mechanics. Presumably this refers to the popular account of this interpretation according to which something can be true in this world but false in another world, so an agent $A$ can be certain that $x = \xi$ while not denying that $x \ne \xi$. But this is equivocation: $x$ in the first statement does not refer to the same thing as $x$ in the second statement. In $A$'s statement that $x = \xi$, the symbol $x$ means ``the result of the measurement in my world'', while in $x\ne\xi$ it means ``the result of the measurement in another world''. In the many-worlds interpretation as much as in any other interpretation, $A$ cannot be certain that $x = \xi$  without denying that $x\ne\xi$ unless they change the meaning of $x$.

This completes the discussion of the assumptions stated by FR. We have found only one version of quantum mechanics that might not satisfy one of these assumptions, namely the case of QBism and Assumption (C). This is summarised in Table 1 (compare Table 4 of \cite{FR2}, which differs in many lines).
 %We must therefore look elsewhere, in the other assumptions not declared by FR, if we are to acquit versions of quantum mechanics of the charge of contradiction.

 %Their acquittal from the charge of contradiction must therefore lie elsewhere, in one of the other assumptions which were not declared by FR.

\begin{table}[ht]
\begin{center}
\begin{tabular}{l|c|c|c|}
& (Q) & (S) & (C) \\
\hline
Copenhagen & $\surd$ & $\surd$ & $\surd$\\
\hline
Collapse theories e.g. \cite{GRW} & $\surd$ & $\surd$ & $\surd$ \\
\hline
Bell-Bohm \cite{QMPN} pp. 215-217 & $\surd$ & $\surd$ & $\surd$ \\
\hline
Relative-state \cite{Everett, Wheeler, QMPN, Rovelli:relational, logicfuture}& $\surd$ & $\surd$ & $\surd$\\
\hline
Many worlds \cite{deWittGraham} & $\surd$ & $\surd$ & \ $\surd$  \\
\hline
Consistent histories \cite{Griffiths:book} & $\surd$ & $\surd$& $\surd$\\
\hline
QBism \cite{QBism}& $\surd$ & $\surd$ & $\times$\\
\hline
\end{tabular}
\end{center}
\caption{The assumptions of FR}
\end{table}

We see that in all cases except for QBism, we must look elsewhere, in the other assumptions not declared by FR, if we are to acquit versions of quantum mechanics of the charge of contradiction.

\subsection{}\label{P}
\upline
\begin{quote} {\bf Assumption (P)}\ \  If an agent $A$ prepares a system $S$ in a state $|\psi\>$, then immediately after the preparation $A$ is certain that $S$ is in the state $|\psi\>$.\end{quote}

This appears to be obvious, and guaranteed by the very meaning of preparation of a state. However, if $F_1$ is an adherent of the ``many worlds'' or ``relative state'' interpretations of quantum mechanics, then they know that they have become entangled with the coin by their measurement in Stage EWF0, and therefore the coin is in a mixed state.  What, then, do they determine when they observe the coin and prepare it? In ``many worlds'' terms, they know that there is a world in which they got the result ``heads'' on observing the coin and prepared it in the state $|\downarrow\>$, but they find out that they are in the other world in which they got the result ``tails'' on observing the coin and prepared it in the state $|\rightarrow\>$. In ``relative state'' terms, they know that the absolute state is the entangled state, but that the state relative to themself is what they observe and prepare. Thus this assumption amounts to ignoring the difference between the absolute state and the relative state.

\subsection{}\label{U}
\upline
\begin{quote} {\bf Assumption (U)}\ \ If an agent $A$ is certain at time $t$ that the external world is in a state $|\Psi\>$ and $A$ knows the Hamiltonian $H$ governing the external world, then $A$ is certain at time $t$ that the state of the external world at time $t + T$ will be $\e^{-iHT/\hbar}|\Psi\>$.\end{quote}

This looks like an essential part of quantum mechanics in any interpretation. However, it is violated in the Copenhagen interpretation (and therefore in most textbook presentations of quantum mechanics), since that requires that unitary evolution is interrupted by the projection that occurs whenever a measurement is made. 

This assumption might also be violated in the Bell-Bohm hidden-variables theory, depending on what the word ``state'' is taken to mean: in that theory there is a pilot state, which satisfies assumption (U), and the actual state, which does not. Likewise in the relative-state interpretation, the meaning of the assumption depends on whether the word ``state'' refers to the external or the internal state. However, the use of this assumption requires ``state'' to refer to the pilot state in the Bell-Bohm theory, and the external state in the relative-state interpretation, so in both these cases we can take it that Assumption (U) is satisfied. These meanings of ``state'', however, render Assumption (P) invalid.

\subsection{}\label{T}
\upline
\begin{quote} {\bf Assumption (T)}\ \  If the possible values of a measurement result are $r_1, r_2, \ldots$ and an agent $A$ is certain that the result was not $r_1$, then $A$ is certain that the result was one of $r_2, \ldots$. \end{quote}

This assumption is satisfied if agent $A$ is adopting the Copenhagen interpretation, where a measurement result is taken to be a classical variable. This requires that all the other agents are also adopting this interpretation, if Assumption (C) is to be applicable. In Bell-Bohm theory it is also satisfied if the actual state is taken to be defined by the memories of all the agents, as in \cite{singleworld}; but other versions of this theory are possible, in which Assumption (T) may not be satisfied.

In the relative-state interpretation assumption (T) is satisfied if it refers to a measurement performed by the agent $A$, who will then treat the result as a classical variable, but not if it was performed by any of the other agents, who, for $A$, are quantum sytems and the result of the measurement is a quantum variable.

\subsection{}\label{L}
\upline
\begin{quote} {\bf Assumption (L)}\ \  If there is no action in the neighbourhood of a system $F$ in a certain time interval, then the state of $F$ remains unchanged during that interval.\end{quote}

This is an assumption of locality, which is well known, following the work of John Bell \cite{Bell:book}, to be violated in hidden-variable interpretations of quantum mechanics. In particular, it is violated by the Bell-Bohm interpretation discussed above. However, the application of this interpretation shown in subsection \ref{C} actually satisfies Assumption (L).

  Long before Bell's work, it was pointed out by Einstein (\cite{Jammer:book} p.373) that the projection postulate of the Copenhagen interpretation also displays a violation of locality. 
 
\subsection{}\label{M}
\upline
\begin{quote} {\bf Assumption (M)}\ \  Memory states of an agent are irreversible and do not change once they have been established unless there is some surgical operation on, or neural pathology in, the agent.\end{quote}

This is satisfied by the Copenhagen interpretation, in which the memory of an agent can be taken as the result of a measurement, though it is doubtful whether the qubit systems which model agents in the FR Gedankenexperiment could be accepted as measurement apparatuses in the spirit of the Copenhagen interpretation. Stages EWF3 and EWF4 of the Gedankenexperiment, in which the agents and their labs are measured in a basis of superpositions of memory states, must be regarded as surgical operations. 

Spontaneous collapse theories would also satisfy this assumption if the agents' memories were macroscopic systems, but not for FR's model of memories as qubit systems. The assumption is not necessarily satisfied by the Bell-Bohm interpretation, though, as for locality, the particular history derived in subsection \ref{C} does satisfy it (the change between EWF3 and EWF4 being taken as the result of surgery).

\subsection{Summary}

The status of all the assumptions in various interpretations and versions of quantum mechanics is shown in the table below. An interpretation must violate one of (Q), (C), (S), (P), (U), (T), or both of (L) and (M), in order to escape FR's charge of inconsistency.

\begin{table}[ht]
\begin{center}
\begin{tabular}{l|c|c|c|c|c|c|c|c}
& (Q) & (S) & (C) & (P) & (U) & (T) & (L) & (M) \\
\hline
Copenhagen & $\surd$ & $\surd$ & $\surd$ & $\surd$ & $\times$ & $\surd$ & $\times$ & $\surd$\\
\hline
Collapse theories e.g. \cite{GRW} & $\surd$ & $\surd$ & $\surd$ & $\surd$ & $\times$ & $\surd$ & $\times$ & $\times$ \\
\hline
Bell-Bohm \cite{QMPN} pp. 215-217 & $\surd$ & $\surd$ & $\surd$ & $\surd$ & $\times$ & $\surd$ & $\times$ & $\times$\\
\hline
Relative-state \cite{Everett, Wheeler, QMPN, Rovelli:relational, logicfuture} & $\surd$ & $\surd$ & $\surd$ & $\times$ &$\surd$ & $\times$ & $\times$ & $\times$\\
\hline
Many worlds \cite{deWittGraham} & $\surd$ & $\surd$ & $\surd$ &$ \times$ & $\surd$ & $\times$ & $\times$ & $\times$ \\
\hline
Consistent histories \cite{Griffiths:book} & $\surd$ & $\surd$ & $\surd$ & $\surd$ & $\surd$ & $\surd$ & $\times$ & $\surd$ \\
\hline
QBism \cite{QBism} & $\surd$ & $\surd$ & $\times$ & $\surd$ & $\surd$ & $\surd$ & $\surd$ & $\surd$ \\
\hline
\end{tabular}
\end{center}
\caption{All assumptions}
\end{table}

I have not argued in detail for all of these attributions; they may be disputed. If valid, however, they show that the consistency property expressed in Assumption (C) does not have to be dropped by any interpretation of quantum mechanics except, possibly, QBism.

\section{Conclusion}

We have examined the argument of Frauchiger and Renner, in which a contradiction is deduced from three assumptions whose joint content can be expressed as ``Quantum mechanics can describe the use of itself''. We have found that a number of other assumptions are made in deducing the contradiction, and that several interpretations of quantum mechanics fail to satisfy one or more of these other assumptions, and can therefore consistently be held without accepting the contention of Frauchiger and Renner that quantum mechanics cannot describe the use of itself.

%\bibliography{quantum}
%\bibliographystyle{plain}

\end{document}